\documentclass{article}
\def\be{\begin{equation}}
 
\def\becenter{\begin{center}}
\def\encenter{\end{center}}
\def\ee{\end{equation}}
\def\delt12{\delta_{12}}

\begin{document}
\title{\bf Tunnel Effect or `Saute-Mouton'?}         

\author{  {\bf Antonio Carlos B. Antunes$^\dag$ and Leila J. Antunes$^\ddag$} \\ 
$\dag$ Instituto de F\'\i sica, Universidade Federal do Rio de
Janeiro\\ C.P. 68528, Ilha do Fund\~ao\\ 
21945-970 Rio de Janeiro, RJ, Brazil \\
e-mail: acbantun@uninet.com.br or antoniocbantunes@gmail.com \\ 
$\ddag$ Instituto de Engenharia Nuclear - CNEN\\C.P. 68550, Ilha do Fund\~ao\\
21945-970 Rio de Janeiro, RJ, Brazil \\e-mail: leila@ien.gov.br}

\date{ }

\maketitle

\abstract{ An infinite well potential containing a rectangular barrier in its center is used to verify if the passage of a quantum particle through the barrier is described by tunnel effect or `saute-mouton'.

} 

\section{Introduction}

In the book {\it Quantique}\cite{quantique}, the authors, J.M. L\'evy-Leblond and F. Balibar, conjecture that the tunnel effect in quantum mechanics can be interpreted as if the quantum particle surmounts the potential barrier. They argue that if the incident particle has an energy E lower than the potential energy $\mathrm{V_b}$ of the barrier, along the interaction the uncertainty $\Delta \mathrm{E}$ in the particle energy provides the amount of energy necessary for the particle to overpass the potential barrier.

In this paper, we consider a simple example in which the energy E and the uncertainty $\Delta \mathrm{E}$ of the quantum system can be calculated in order to be compared with the potential energy $\mathrm{V_b}$ of the barrier.  This example consists of an infinite well containing a rectangular barrier in its middle. The potential energy is defined by
\begin{eqnarray}
V(x) &=& \mathrm{V_b} \enspace \mathrm{for} \enspace | \thinspace x \thinspace | < b \thinspace , \nonumber \\ 
V(x) &=& 0 \enspace \mathrm{for}  \enspace b < | \thinspace x \thinspace | <a \thinspace , \mathrm{and}  \nonumber \\
V(x) &=& +\infty \enspace \mathrm{for}  \enspace | \thinspace x \thinspace | > a \thinspace .\nonumber
\end{eqnarray}
We solve the Schr\"odinger equation for $\mathrm{E} <\mathrm{V_b}$ and choose units in which $\hbar = 1 $ and $2\mathrm{m} = 1$.

Using the definitions $\mathrm{k}^2 = \mathrm{E}$ and $\mathrm{K}^2 = \mathrm{V_b} - \mathrm{E}$ \thinspace, the Schr\"odinger equation becomes
\begin{eqnarray}
\frac{\mathrm{d}^2 \psi}{\mathrm{d x}^2} + \mathrm{k}^2 \psi &= &0 \enspace \mathrm{for} \enspace \mathrm{b} < \mathrm{| \thinspace x \thinspace |} < \mathrm{a} \thinspace , \quad \mathrm{and} \nonumber \\
\frac{\mathrm{d}^2 \psi}{\mathrm{d x}^2} - \mathrm{K}^2 \psi &=& 0 \enspace \mathrm{for} \enspace  \mathrm{| \thinspace x \thinspace |} < \mathrm{b} \thinspace .\nonumber
\end{eqnarray}

Solving these equations and imposing the boundary conditions 
\[
\psi( \pm \mathrm{a} ) = 0
\]
and the continuity conditions
\begin{eqnarray}
\psi (\mathrm {x - \epsilon}) &=& \psi (\mathrm{x + \epsilon}) \quad \mathrm{and} \nonumber \\
\psi' (\mathrm {x - \epsilon}) &=& \psi' (\mathrm{x + \epsilon})  \nonumber
\end{eqnarray}
for $\mathrm{x} = \pm \thinspace \mathrm{b}$ and $\epsilon \rightarrow 0 \thinspace$, we obtain the eigenvalue equations.

For the symmetric and antisymmetric solutions the eigenvalues are given by\cite{cohen}
\begin{eqnarray}
\frac{ \mathrm{cotanh(K b)} } {\mathrm{K}} + \frac{\mathrm{tan}[\thinspace \mathrm{k ( a - b)}\thinspace]} {\mathrm{k}}&=& 0 \quad \mathrm{and} \nonumber \\
\frac{\mathrm{tanh(K b)}} {\mathrm{K}} +  \frac{\mathrm{tan}[\thinspace\mathrm{k ( a - b)}\thinspace]} {\mathrm{k}}&=& 0 \quad \mathrm{respectively.} \nonumber
\end{eqnarray}

\section{Time dependent solution.}
Let $\mathrm{E_S}$ and $\mathrm{E_A}$ be the lowest energy eigenvalues for the symmetric and antisymmetric solutions respectively, and let $\psi_{\mathrm{S}}(\mathrm{x})$ and $\psi_{\mathrm{A}}(\mathrm{x})$ denote the normalized eigenfunctions of these states.
The time dependent solution is\cite{merz}
\[
\psi(\mathrm{x,t}) = \frac{1}{\sqrt{2}} \bigl\{  \psi_{\mathrm{S}}(\mathrm{x}) e^{-  \mathrm{i \thinspace E_S \thinspace \mathrm{t}}} +  \psi_{\mathrm{A}}(\mathrm{x}) e^{-  \mathrm{i \thinspace  E_A \thinspace \mathrm{t}} }  \bigr\}
\] 
which can be rewritten as 
\[
\psi(\mathrm{x,t}) = e^{- \mathrm {i \thinspace E \thinspace \mathrm{t}} } \biggl{\{} \psi_{\negthinspace _+}\negthinspace{(\mathrm{x})} \thinspace \mathrm{cos} \Bigl( \frac{\Delta \mathrm{E}} {2} \thinspace \mathrm{t} \Bigr)  +  
\mathrm{i} \thinspace \psi_{\negthinspace _-}\negthinspace{(\mathrm{x})}  \thinspace \mathrm{sen} \Bigl( \frac{\Delta \mathrm{E}} {2} \thinspace \mathrm{t} \Bigr)  \biggr{\}} \thinspace ,
\]
where
\begin{eqnarray}
\mathrm{E} &=& \frac{1}{2} \left( \mathrm{E_A} + \mathrm{E_S}\right) \thinspace , \nonumber \\
\Delta \mathrm{E} &=&  \left(\mathrm{E_A} - \mathrm{E_S} \right) \thinspace , \nonumber \\ 
\psi_{\negthinspace _+} &=& \frac{1}{\sqrt{2}} \bigl( \psi_{\mathrm{S}} + \psi_{\mathrm{A}} \bigr) \quad \mathrm{and} \nonumber \\
\psi_{\negthinspace _-}&=& \frac{1}{\sqrt{2}} \bigl( \psi_{\mathrm{S}} - \psi_{\mathrm{A}} \bigr)\thinspace . \nonumber
\end{eqnarray}

From the time solution we obtain the oscillating density of probability
\[
|\thinspace\psi{(\mathrm{x,t})} |^2 = |\thinspace \psi_{\negthinspace _+}\negthinspace{(\mathrm{x})}  |^2 \mathrm{cos}^2 \Bigr( \frac{\Delta \mathrm{E}}{2}\thinspace  \mathrm{t} \Bigr) + |\thinspace \psi_{\negthinspace _-}\negthinspace{(\mathrm{x})}  |^2 \mathrm{sen}^2 \Bigr( \frac{\Delta \mathrm{E}}{2}\thinspace  \mathrm{t} \Bigr) \thinspace .
\]

The period and the angular frequence of the oscillations are :
\[
\mathrm{T} = \frac{2 \pi}{\Delta \mathrm{E}} \quad \mathrm{and} \quad  \omega = \Delta \mathrm{E} \thinspace .
\]
\section{Numerical results.}

We chose some convenient values for the parameters a, b and $\mathrm{V_b}$, in order to verify if there exist counterexamples to the conditions required for the occurrence of the `saute mouton'.  Table 1 shows the values of the energy E, uncertainty $\Delta \mathrm{E}$ and the period of oscillation T for a=1.0 and several values of b and $\mathrm{V_b}$.
The results in this table show that in these examples the condition $ \mathrm{V_b} < \mathrm{E}\ \thinspace + \thinspace  \Delta \mathrm{E} $ is not satisfied. We conclude that these are valid counterexamples to the surmount of the potential barrier and the correct interpretation is the tunneling.

\begin{table}[h]
\centering 
\caption{Energy E, uncertainty $\Delta \mathrm{E}$, period of oscillation T}%
 for a=1.0 and different b and $\mathrm{V_b}$ values.  
\begin{tabular}{c c c c c c} 
\\ 
\hline 
b & $ \mathrm{V_b}$ & $\mathrm{E}$ & $\Delta \mathrm{E}$ & $\mathrm{T}$ & $ \mathrm{V_b} > \mathrm{E}\ \thinspace + \thinspace \mathrm{n} \thinspace \Delta \mathrm{E} $ \\ [0.5ex] 
\hline 
1./3.& 15. &10.792&2.819 & 2.229 &$\mathrm{V_b} > \mathrm{E}\ \thinspace + \thinspace  \Delta \mathrm{E} \thinspace \thinspace \thinspace \thinspace \thinspace  $ \\
0.2 & 25. & 9.642 & 2.387 & 2.632 &$\mathrm{V_b} > \mathrm{E}\ \thinspace + \thinspace \thinspace 6 \thinspace \Delta \mathrm{E}$  \\ 
0.1 & 50. & 8.931 & 2.429 & 2.586 &$\mathrm{V_b} > \mathrm{E}\ \thinspace + \thinspace 16 \thinspace \Delta \mathrm{E}$  \\ [1ex]
\hline 
\end{tabular}
\label{table:tableone} 
\end{table}

Clearly, the results above depend on the definition of the uncertainty in energy $\Delta \mathrm{E}$. 

In a rigorous derivation of the uncertainty principle for energy and time, Mandelstam and Tamm\cite{mandelstam} define the energy uncertainty in terms of the standard deviation
\[
\sigma_{_\mathrm{E}} = \sqrt{\langle \thinspace(\mathrm{H} - \langle \mathrm{H} \rangle )^2 \rangle } \thinspace \thinspace  ,
\]
where
\[
\langle \mathrm{H} \rangle = \langle \psi(x,t) |\thinspace  \mathrm{H}\thinspace | \psi(x,t) \rangle \thinspace \thinspace  ,
\]
which gives
\[
\sigma_{_\mathrm{E}} = \frac{1}{2} ( \mathrm{ E_A - E_S} ) \thinspace \thinspace .
\]
Thus, this definition of uncertainty reinforces the argument in favor of tunneling.

{}
\end{document}